\documentclass[10pt, conference]{IEEEtran}

\RequirePackage[normalem]{ulem} 
\RequirePackage{color}\definecolor{RED}{rgb}{1,0,0}\definecolor{BLUE}{rgb}{0,0,1} 

\RequirePackage{color}

\usepackage{color}%
\usepackage{cite}
\usepackage{amsthm}
\usepackage[cmex10]{amsmath}
\usepackage{array}
\usepackage{mdwmath}
\usepackage{mdwtab}
\usepackage{amsfonts}
\usepackage{amssymb}
\usepackage{eqparbox}
\usepackage{multirow}
\usepackage{graphicx}
\usepackage[labelsep=period]{caption}
\usepackage{caption}
\usepackage{subcaption}
\usepackage{graphicx}
\usepackage{color}
\usepackage{multirow}

\usepackage{algorithm}
\usepackage{algorithmic}

\usepackage[belowskip=2pt,aboveskip=0pt]{caption}
\newcommand{\argmin}{\operatornamewithlimits{argmin}}
\begin{document}
\title{High-Rate Space Coding for Reconfigurable $2\times 2$ Millimeter-Wave MIMO Systems}
\author{\IEEEauthorblockN{Vida Vakilian\IEEEauthorrefmark{1}\thanks{This work was in part supported by Department of Defense Grant, number W911NF-15-1-0033.}, 
Hani Mehrpouyan\IEEEauthorrefmark{2}, Yingbo Hua\IEEEauthorrefmark{1}, and Hamid Jafarkhani\IEEEauthorrefmark{3} } \\

\IEEEauthorblockA{\IEEEauthorrefmark{2}  Dept. of Elect. and Comp. Eng. and Comp. Science, California State University, Bakersfield, USA, hani.mehr@ieee.org} 
\IEEEauthorblockA{\IEEEauthorrefmark{1} Dept. of Elect. Eng., University of California, Riverside, USA, vida.vakilian.ca@ieee.org, yhua@ee.ucr.edu}
\IEEEauthorblockA{\IEEEauthorrefmark{3} Center for Pervasive Communication \& Computing, University of California, Irvine, USA, hamidj@uci.edu}
}
\maketitle
\begin{abstract}
Millimeter-wave links are of a line-of-sight nature. Hence, multiple-input multiple-output (MIMO) systems operating in the millimeter-wave band may not achieve full spatial diversity or multiplexing. In this paper, we utilize \textit{reconfigurable antennas} and the high antenna directivity in the millimeter-wave band to propose a \textit{rate-two} space coding design for $2\times2$ MIMO systems. The proposed scheme can be decoded with a low complexity maximum-likelihood detector at the receiver and yet it can enhance the \textit{bit-error-rate} performance of millimeter-wave systems compared to traditional spatial multiplexing schemes, such as the Vertical Bell Laboratories Layered Space-Time Architecture (VBLAST).
Using numerical simulations, we demonstrate the efficiency of the proposed code and show its superiority compared to existing \textit{rate-two} space-time block codes.

\end{abstract}
\section{Introduction}
The millimeter-wave (mmWave) technology operating at frequencies in the $30$ and $300$ GHz range is considered as a potential solution for the $5$th generation (5G) wireless communication systems to support multiple gigabit per second wireless links~\cite{daniels200760,rappaport2013millimeter}. The large communication bandwidth at mmWave frequencies will enable mmWave systems to support higher data rates compared to microwave-band wireless systems that have access to very limited bandwidth. However, significant pathloss and hardware limitations are major obstacles to the deployment of mm-wave systems.

In order to combat their relatively high pathloss compared to systems at lower frequencies and the additional losses due to rain and oxygen absorption, mmWave systems require a large directional gain and line-of-sight (LoS) links. This large directional gain can be achieved by beamforming either using a large antenna array or a \textit{single} reconfigurable antenna element, which has the capability of forming its beam electronically\cite{cetiner2004multifunctional,cetiner2006mimo,
vakilian2012performance,piazza2008design,
frigon2008dynamic,vakilian2014covariance,li2009capacity,vakilian2013space}. Such reconfigurable antennas are available for commercial applications. As an example, composite right-left handed (CRLH) leaky-wave antennas (LWAs) are a family of reconfigurable antennas with those characteristics \cite{lim2004electronically,caloz2008crlh}.
By employing reconfigurable antenna elements where \textit{each} antenna is capable of configuring its radiation pattern independent of the other antennas in the array, a LoS millimeter-wave multiple-input multiple-output (MIMO) system can achieve both multiplexing and diversify gains. The former will result in better utilization of the bandwidth in this band, while the latter can allow designers to overcome the severe pathloss~\cite{sayeed2013beamspace}.

Although the advantages of reconfigurable antennas are well-documented, the space coding designs for MIMO systems are mostly considered based on the assumption that the antenna arrays at the transmitter and the receiver are omni-directional, i.e., there is no control mechanism over the signal propagation from \textit{each} antenna element~\cite{foschini1996layered,alamouti1998simple,tarokh1999space}. Deploying reconfigurable antennas in MIMO arrays can add multiple degrees of freedom to the system that can be exploited to design new space coding designs that improve the system performance compared to existing schemes. 

In recent years, several block-coding techniques have been designed to improve the performance of MIMO systems employing reconfigurable antennas. In~\cite{grau2008reconfigurable}, the authors have proposed a coding scheme that can increase the diversity order of conventional MIMO systems by the number of the reconfigurable states at the receiver antenna. \cite{fazel2009space} extends the technique in~\cite{grau2008reconfigurable} to MIMO systems with reconfigurable antenna elements at both the transmitter and receiver sides, where a state-switching transmission scheme is used to further utilize the available diversity in the system over flat fading wireless channels. However, using the coding schemes introduced in \cite{grau2008reconfigurable} and \cite{fazel2009space} the system is only able to transmit one symbol per channel use, i.e., they do not provide any multiplexing gain. Moreover, the detection complexity of the codes in~\cite{grau2008reconfigurable} and \cite{fazel2009space} is high and increases with the number reconfigurable states at the antenna.

\vspace{-2pt}
\subsection{Contributions}

In this paper, we propose a rate-two space encoder for $2 \times 2$ MIMO systems equipped with reconfigurable transmit antennas. The proposed encoder uses the properties of reconfigurable antennas to achieve multiplexing gain, while reducing the complexity of the maximum-likelihood (ML) detector at the receiver. Compared to previously proposed space coding schemes outlined below, the proposed design utilizes the reconfigurability of the antennas to increase bandwidth efficiency, enhance reliability, and reduce detection complexity at the receiver. In fact, the proposed encoder has a detection complexity of ${\cal O}(M)$, where $M$ is the cardinality of the signal constellation. These advantages are made possible since we have utilized the high antenna directivity at mmWave frequencies~\cite{art-hani-eband-14} and the reconfigurability of the antennas to ensure that the beams from each reconfigurable antenna is directed at a receive antenna as shown in Fig.~1. Hence, in a $2\times 2$ MIMO system, the proposed approach can generate four beams for each transmit-to-receive antenna pair that can be modified via the reconfigurable antenna parameters. On the other, the conventional MIMO beamforming scheme for omnidirectional antennas is only capable of generating a maximum of two beams in a similar setup \cite{zheng2007mimo}. Thus, although effective, the conventional MIMO beamforming cannot be applied to circumvent ill-conditioned LoS MIMO channels in the mm-wave band.


For comparison purposes, we compare the performance of the proposed encoder against the Vertical Bell Laboratories Layered Space-Time Architecture (VBLAST)~\cite{foschini1996layered} for detection via successive interference cancellation (SIC) and ML. The results of our investigation demonstrate that the proposed approach can outperform SIC- and ML-VBLAST, while requiring the same decoding complexity at the receiver as SIC-VBLAST. We also study the performance of the recently developed rate-2 space-time block codes (STBCs), including the \textit{Matrix C} \cite{IEEEStd2006Air}, and \textit{maximum transmit diversity (MTD)}~\cite{rabiei2009new} codes. The Matrix C code is a threaded algebraic space-time code \cite{damen2003linear}, which is known as one of the well-performing STBCs for $2 \times 2$ MIMO systems. 
However, the ML decoding complexity of this code is very high; it is ${\cal O}(M^4)$, i.e., an order of four. Similarly, the MTD code~\cite{rabiei2009new} has an ML detection complexity of ${\cal O}(M^2)$. Although a rate-2 STBC for MIMO systems equipped with reconfigurable antennas is proposed  in~\cite{vakilian2015STBC}, the detection complexity of the proposed code is of an order of ${\cal O}(M^2)$. Furthermore, the STBC in~\cite{vakilian2015STBC} is designed based on the assumption that the radiation pattern of each reconfigurable antenna consists of a single main lope with negligible side lopes. Thus, by not utilizing the side lopes, the higher detection complexity of the code in~\cite{vakilian2015STBC} does not translate into better overall system performance.


\vspace{-2pt}
\subsection{Organization}
The rest of the paper is organized as follows. In Section~\ref{sec:sys_model}, we describe the system and signal model. In Section~\ref{sec:CodeConstruction}, we introduce the proposed high-rate code for $2 \times 2$ MIMO systems. We describe the design criteria of the code in Section~\ref{sec:designcriteria}. We present a low complexity ML decoder for the proposed code in Section~\ref{sec:Decoding}. Simulation results are presented in Section~\ref{sec:results}, and concluding remarks are provided in Section~\ref{sec:conc}.

\vspace{-2pt}
\subsection{Notation}
Throughout this paper, we use capital boldface letters, $\mathbf{X}$, for matrices and lowercase boldface letters, $\mathbf{x}$, for vectors. $(\cdot)^T$ denotes transpose operator. ${\bf A}\circ {\bf B}$ denotes the Hadamard product of the matrices ${\bf A}$ and ${\bf B}$, $||{\bf A}||_F$ represents the Frobenius norm of the matrix ${\bf A}$, $\text{det}({\bf A})$ computes the determinant of the matrix ${\bf A}$, and vec({\bf A}) denotes the vectorization of a matrix {\bf A} by stacking its columns on top of one another. Moreover, $\text{diag}(a_1, a_2, \cdots, a_n)$ represents a diagonal $n \times n$ matrix, whose diagonal entries are $a_1, a_2, \cdots , a_n$. ${\bf I}_M$ denotes the identity matrix of size $M\times M$. Finally, $\mathbb{C}$ denotes the set of complex valued numbers.



\section{System Model and Definitions}
\begin{figure}
\centering
\includegraphics [scale=0.65]{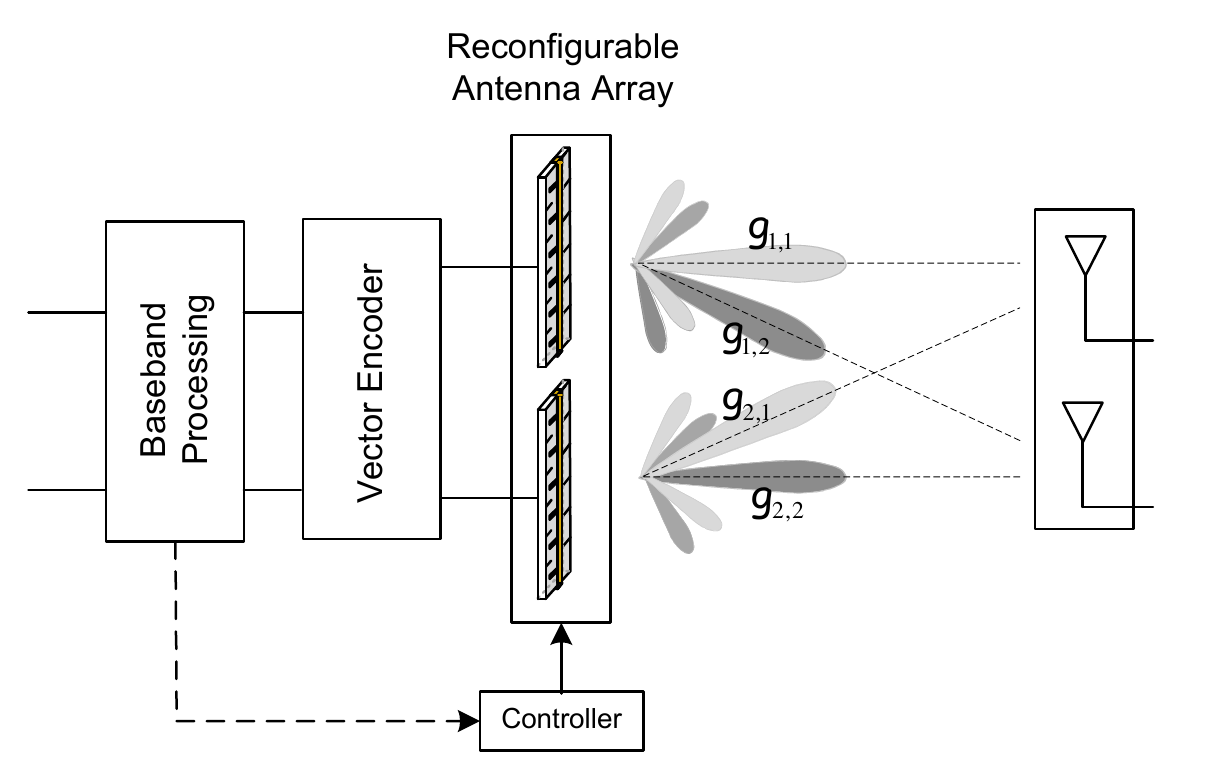}
\caption{\small Reconfigurable MIMO system transmitter.}
\label{fig:system_model}
\end{figure}
\label{sec:sys_model}
We consider a MIMO system with $N_t = 2$ transmit and $N_r = 2$ receive antennas. The transmit antennas are assumed to be reconfigurable with controllable radiation patterns \cite{li2013new} and the receive antennas are assumed to be omni-directional, see Fig.~\ref{fig:system_model}. Due to the utilization of the mmWave band, we assume that the wireless channels between each pair of the transmit and receive antennas are Rician flat fading~\cite{art-hani-eband-14}.
Based on the above assumptions, the received signal can be expressed as
\begin{align}
{\bf y} = {\bf H}_g {\bf c} + {\bf z},
\label{eq:receivedsignalmatrix}
\end{align}
where ${\bf c} = [c_{\!_1}, c_{\!_2}, \cdots, c_{\!_{N_t}}]^T \in \mathbb{C}^{N_t \times 1}$ is the transmitted code vector, ${\bf z} \in \mathbb{C}^{N_r \times 1}$ is a zero-mean complex white Gaussian noise matrix consisting of components with identical power $N_0$, and ${\bf H}_g \in \mathbb{C}^{N_r \times N_t}$ is the Hadamard product of the channel matrix ${\bf H}$ and the reconfigurable antenna parameter matrix ${\bf G}$, i.e.,
\begin{align}
{\bf H}_{g} = {\bf H} \circ {\bf G}.
\label{eq:Hg}
\end{align}
In (\ref{eq:Hg}), ${\bf H} \triangleq \big[{\bf h}_{1}, \cdots, {\bf h}_{N_t} \big]$ with ${\bf h}_j \triangleq [h_{1,j}, \cdots, h_{N_r,j}]^T, \label{eq:ch}$ and
${\bf G} \triangleq \big[{\bf g}_1, \cdots, {\bf g}_{\!{N_t}}\big]$ with ${\bf g}_j \triangleq [g_{\!_{1,j}}, \cdots, g_{\!_{N_r,j}}]^T$. Here, $h_{i,j}$ and $g_{\!_{i,j}}$ denote the channel and reconfigurable antenna parameters corresponding to the $i$th and $j$th receive and transmit antennas, respectively.

Note that since the radiation pattern towards each receive antenna can be modified independent of the other antennas, a Hadamard product instead of a general vector multiplication is used in~\eqref{eq:Hg}.

\textit{Definition 1:}({\bf Transmission rate}) If $N_s$ information symbols in a codeword are transmitted over $T$ channel uses,  the transmission symbol rate is defined as
\begin{align}
\nonumber
r_s  = \frac{N_s}{T},
\end{align}
and the bit rate per channel use is then given by
\begin{align}
\nonumber
r_b  = r_s \log_2 M,
\end{align}
where $M$ is the cardinality of the signal constellation. \vspace{5pt}

\textit{Definition 2:}({\bf Maximum-likelihood decoding complexity}) The maximum-likelihood decoding metric that is to be minimized over all possible values of codeword ${\bf c}$ is given by
\begin{align}
 [{\hat c}_1, \cdots, {\hat c}_{N_t}]={\argmin_{c_1, \cdots, c_{N_t}}} \,||{\bf y}-{\bf H}_g{\bf c}||^2.
 \label{eq:joint-ML_total}
\end{align}
If we assume that there are $N_s$ symbols to be transmitted in each codeword, then the ML decoder complexity  will be ${\cal O}(M^{N_s})$ for joint data detection. As we will show in the sequel, we can reduce the ML complexity of the proposed code to ${\cal O}(M)$ using the structure of the code and the reconfigurable feature of the antennas.

\section{Code Construction}
\label{sec:CodeConstruction}
Let us consider a $2\times2$ MIMO system.  We construct every $2\times 1$  codeword vector from two information symbols  $\{s_1, s_2\}$  that will be sent from $N_t = 2$ reconfigurable antennas. The proposed codeword ${\bf c}$ can be expressed as
\begin{align}
{\bf c} &= \frac{1}{\sqrt{\nu}}\left[
\begin{array}{cc}
   \alpha_1  &  \beta_1     \\
   \alpha_2 &  \beta_2
\end{array}
\right] {\bf s},
\nonumber
\end{align}
where ${\bf s} = [s_1, s_2]^T$ is transmit symbol vector. Therefore, the codeword ${\bf c}$ is given by
\begin{align}
{\bf c} = \frac{1}{\sqrt{\nu}}\left[
\begin{array}{c}
   \alpha_1 s_1 +  \beta_1 s_2   \\
   \alpha_2 s_1 + \beta_2 s_2
   \end{array}
\right],
\label{eq:2x2BlockCode}
\end{align}
where $\nu$ is the power normalization factor and $\alpha_1$, $\beta_1$, $\alpha_2$ and $\beta_2$ are design parameters that are chosen to provide the maximum diversity and coding gain.



\section{Design Criteria}
\label{sec:designcriteria}
In this section, we first discuss the diversity order of the proposed code and then discuss mechanism for obtaining the optimal values for $\alpha_1$, $\beta_1$, $\alpha_2$ and $\beta_2$.

To compute the achievable diversity gain of the proposed code, consider two distinct codewords ${\bf c}$ and ${\bf u}$ that are constructed using (\ref{eq:2x2BlockCode}) as
\setcounter{equation}{3}
\begin{subequations}
\nonumber
\begin{align}
{\bf c} &= \frac{1}{\sqrt{\nu}}\left[
\begin{array}{cc}
   \alpha_1 s_1 +  \beta_1 s_2  \\
   \alpha_2 s_1 + \beta_2 s_2
\end{array}
\right],
\label{eq:2x2BlockCode1}\\
{\bf u} &= \frac{1}{\sqrt{\nu}}\left[
\begin{array}{cc}
   \alpha_1 u_1 +  \beta_1 u_2    \\
   \alpha_2 u_1 + \beta_2 u_2
\end{array}
\right].
\label{eq:2x2BlockCode2}
\end{align}
\end{subequations}
The pairwise error probability (PEP) of the proposed code can be expressed as
\begin{equation}
P(\boldsymbol{\cal C}\rightarrow \boldsymbol{\cal U}|{\bf h}_g)= Q
\Big(\sqrt{\frac{\gamma}{4} ||({\bf
\boldsymbol{\cal C}}-{\bf \boldsymbol{\cal U}}) {\bf h}_g||^2}\Big), \label{eq:PEP}
\end{equation}
where ${\bf \boldsymbol{\cal C}} = ({\bf I}_2 \otimes {\bf c}^T)$, ${\bf \boldsymbol{\cal U}} = ({\bf I}_2 \otimes {\bf u}^T)$, ${\bf h}_g = \text{vec}({\bf H}_g)$, and $\gamma$ is the received signal-to-noise ratio (SNR). By applying the Chernoff upper bound, $Q(x)\leq1/2e^{-x^2/2}$, and calculating the expected value of the upper bound, the pairwise error probability for the proposed code can be upper-bounded by
\begin{equation}
\nonumber
P(\boldsymbol{\cal C}\rightarrow \boldsymbol{\cal U}) \leq \frac{1}{\text{det}\big({\bf I}_4+(\gamma/4)({\bf R}_{h_g}({\bf
\boldsymbol{\cal C}}-{\bf \boldsymbol{\cal U}})^H ({\bf \boldsymbol{\cal C}}-{\bf \boldsymbol{\cal U}})\big)},
\label{eq:PEP}
\end{equation}
where ${\bf R}_{h_g} = {\mathbb E}\{{\bf h}_g{\bf h}_g^H\}$. At high SNR, the above equation can be simplified to
\begin{equation}
P(\boldsymbol{\cal C}\rightarrow \boldsymbol{\cal U}) \leq \frac{1}{(\gamma/4)^r \Pi^{r}_{i=1} \lambda_i},
\label{eq:PEP}
\end{equation}
where $\lambda_i$ and $r$ are the $i$-th eigenvalue and the rank of the matrix ${\bf R}_{{h}_g}({\bf \boldsymbol{\cal C}}-{\bf \boldsymbol{\cal U}})^H ({\bf \boldsymbol{\cal C}}-{\bf \boldsymbol{\cal U}})$, respectively. In other words, $r$ denotes the diversity gain of the proposed code, which can be at maximum $N_r=2$.


To find the parameters of the reconfigurable antennas and that of the codes, we rewrite the received signal equation in (\ref{eq:receivedsignalmatrix}) as
\begin{align}
\nonumber
{\bf y}={\boldsymbol {\cal H}}_g {\bf s}+{\bf z},
\end{align}
where
\begin{align}
\nonumber
{\boldsymbol {\cal H}}_g \triangleq  \left[
\begin{array}{cc}
   \alpha_1 h_{1,1}g_{1,1}+ \alpha_2 h_{1,2}g_{1,2}  &  \beta_1 h_{1,1}g_{1,1} + \beta_2 h_{1,2}g_{1,2}    \\
   \alpha_1 h_{2,1}g_{2,1}+ \alpha_2 h_{2,2}g_{2,2}  &  \beta_1 h_{2,1}g_{2,1} + \beta_2 h_{2,2}g_{2,2}
\end{array}
\right].
\end{align}
We assume that the channel state information (CSI) is known at the transmitter. In time-division-duplex (TDD) systems, the CSI of the uplink can be used as the CSI for the downlink due to channel reciprocity \cite{marzetta2006fast}. In such a setup, no receiver feedback is required. In order to achieve full diversity the matrix ${\boldsymbol {\cal H}}_g$ must be full rank or equivalently its determinant must be nonzero. This condition may not be satisfied for MIMO mmWave systems due to the LoS nature of the link. However, using reconfigurable antennas and through beam steering one can ensure that the determinant of ${\boldsymbol {\cal H}}_g$\textemdash the equivalent channel matrix for the considered reconfigurable $2\times2$ MIMO system\textemdash is nonzero.

The determinant of ${\boldsymbol {\cal H}}_g$ for a $2 \times 2$ MIMO system is given by
\begin{align}
\text{det}({\boldsymbol {\cal H}}_g)& = \Big(\alpha_1\beta_2-\alpha_2\beta1\Big)\Big(h_{1,1}g_{1,1}h_{2,2}g_{2,2}\nonumber \\
 &\hspace{105pt} -h_{1,2}g_{1,2}h_{2,1}g_{2,1}\Big).
 \label{eq:det_Hg}
\end{align}
The constraint $\text{det}({\boldsymbol {\cal H}}_g) \neq 0$ leads to the following two constraints 
\begin{subequations}
\begin{align}
&\big(\alpha_1\beta_2-\alpha_2\beta1\big) \neq 0,  \label{eq:alpha_beta} \\
&\big(h_{1,1}g_{1,1}h_{2,2}g_{2,2}- h_{1,2}g_{1,2}h_{2,1}g_{2,1}\big) \neq 0.
\label{eq:det_Hg1}
\end{align}
\end{subequations}
\begin{figure}
\centering
\includegraphics [scale=0.47]{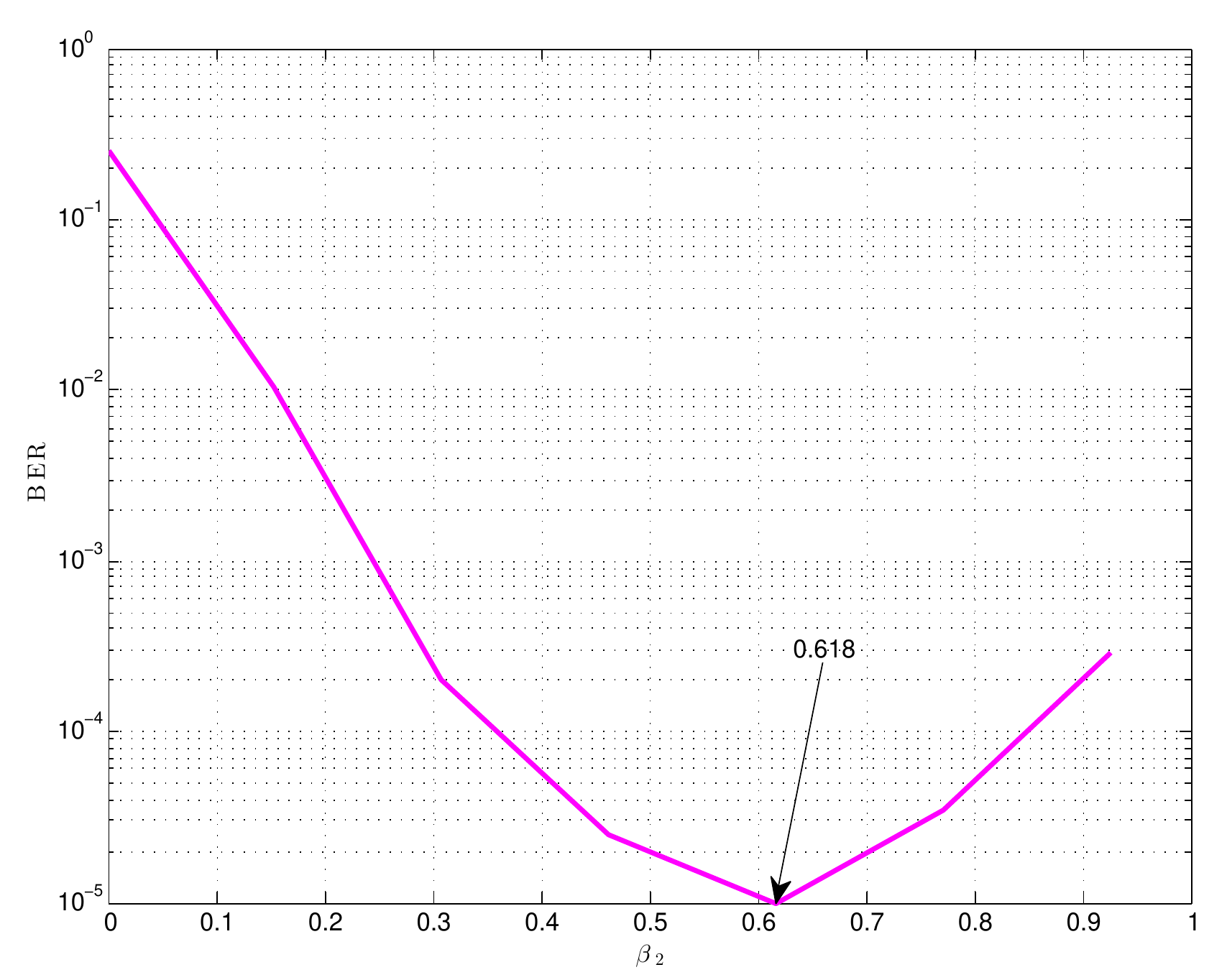}
\caption{\small BER vs. $\beta_2$ for 4-QAM modulation.}
\label{fig:ber-beta2}
\end{figure}
\noindent
For (\ref{eq:alpha_beta}) to be nonzero, we must have
\begin{align}
\alpha_1\beta_2 \neq  \alpha_2\beta1.
\label{eq:kij}
\end{align}
In addition, to control and limit the transmit power of the antennas, the following constraint must be satisfied
\begin{align}
|\alpha_1|^2 + |\beta_1|^2 = |\alpha_2|^2 + |\beta_2|^2 = \nu.
\label{eq:kij2}
\end{align}
Without loss of generality, we set $\alpha_1 = \alpha_2 = 1$. From (\ref{eq:kij}), and (\ref{eq:kij2}), we obtain
\begin{equation}
\nonumber
\beta_1 = -\j \beta_2.
\end{equation}
where $\j= \sqrt{-1}$ is the imaginary unit. We now can find $\beta_2$ analytically by expressing the BER of the system in terms of $\beta_2$ and minimizing it over this parameter. We can also compute this parameter using numerical simulations for a given SNR value. In this paper using the numerical approach we have obtain that $\beta_2 = 0.618$ for 4-QAM signaling at an SNR of $20$ dB, see Fig.~\ref{fig:ber-beta2}.

The parameters of the reconfigurable antennas at the transmitter must be chosen to satisfy~\eqref{eq:det_Hg1} and to reduce the effects of channel fading. As such, the parameters, $g_{i,j}$ for $i,j=\{1, 2\}$, are selected as
\begin{subequations}
\label{eq:antgain}
\begin{align}
g_{1,j} &=  \, h_{1,j}^*/(|h_{1,1}|^2+|h_{1,2}|^2),  \quad \label{eq:antgain1} \\
g_{2,j} &=  \, (-1)^{j}h_{2,j}^*/(|h_{2,1}|^2+|h_{2,2}|^2). \label{eq:antgain2}
\end{align}
\end{subequations}
It can be straightforwardly shown that due to the choice of reconfigurable antenna parameters in (\ref{eq:antgain}), constraint (\ref{eq:det_Hg1}) is satisfied even when the channel matrix, $\mathbf{H}$, is not full-rank due to the LoS nature of the mmWave links.

\section{Decoding}
\label{sec:Decoding}
%
The ML decoder in general performs an exhaustive search over all possible values of the transmitted symbols and decides in favor of the quadruplet ($s_1, s_2$) that minimizes the Euclidean distance metric of (\ref{eq:joint-ML_total}) for a $2\times 2$ system. The computational complexity of the receiver in this case is ${\cal O}(M^2)$. As we will show in the following, the ML decoding complexity of the proposed code can be further decreased to ${\cal O}(M)$.
\subsection{Conditional ML Decoding}
\label{sec:MLConDecoding}
To reduce the decoding complexity of the proposed code, we use a conditional ML decoding technique \cite{sezginer2007full} as elaborated below.
Note that, we set $\alpha_1 = \alpha_2 = 1$ as explained in the previous section. Let us compute the following intermediate signals using the received signals, $y_1$ and $y_2$ as 
\begin{align}
\left\{
\begin{array}{l l}
    y_1= \frac{1}{\sqrt{\nu}} h_{1,1}\,g_{1,1}\,(s_1 + \beta_1 s_2) + \frac{1}{\sqrt{\nu}} h_{1,2}\,g_{1,2}(1)\,(s_1 + \beta_2 s_2 ) + z_1, \label{eq:rxsignal11_2}\\
    y_2 = \frac{1}{\sqrt{\nu}} h_{2,1}\,g_{2,1}\,(s_1 + \beta_1 s_2 ) + \frac{1}{\sqrt{\nu}} h_{2,2}\,g_{2,2}(1)\,(s_1 + \beta_2 s_2 ) + z_2,
   \end{array}
\right..
\end{align}
For a given value of the symbol $s_2$

\begin{align}
r_1 &= y_1 -  \frac{1}{\sqrt{\nu}} \Big(h_{1,1}\,g_{1,1}\, \beta_1 s_2 + h_{1,2}\,g_{1,2}\, \beta_2 s_2 \Big) \nonumber\\
           &=  \frac{1}{\sqrt{\nu}}\Big(h_{1,1}\,g_{1,1}+ h_{1,2}\,g_{1,2}\Big)\,s_1+z_1
\label{eq:r_1_1} 
\\
r_2 &= y_2 -  \frac{1}{\sqrt{\nu}} \Big(h_{2,1}\,g_{2,1}\, \beta_1 s_2 + h_{2,2}\,g_{2,2}\,\beta_2 s_2\Big) \nonumber\\
           &=  \frac{1}{\sqrt{\nu}}\Big(h_{2,1}\,g_{2,1} + h_{2,2}\,g_{2,2}\Big)\,s_1+z_2.
\label{eq:r_2_1}
\end{align}
Now, we form the intermediate signal, ${\tilde r}= r_1 +r_2$, as
\begin{align}
{\tilde r}= \frac{1}{\sqrt{\nu}} &\Big(h_{1,1}\,g_{1,1} + h_{1,2}\,g_{1,2}\nonumber \\&+ h_{2,1} g_{2,1} + h_{2,2}\,g_{2,2}\Big) s_1+{\tilde z},
\label{eq:r_1}
\end{align}
where ${\tilde z} = z_1+z_2$ is the combined noise term. By plugging (\ref{eq:antgain1}) and (\ref{eq:antgain2})  in (\ref{eq:r_1}), we arrive at:
\begin{align}
{\tilde r} =  \frac{1}{\sqrt{\nu}} \Bigg( \sqrt{|h_{1,1}|^2+|h_{1,2}|^2}+ \sqrt{|h_{2,2}|^2-|h_{2,1}|^2}\Bigg)s_1+{\tilde z},
\label{eq:r1}
\end{align}


It can be seen from (\ref{eq:r1}) that ${\tilde r}$ has only terms involving the symbol $s_1$ and, therefore, it can be used as the input signal to a threshold detector to get the ML estimate of the symbol $s_1$ conditional on $s_2$. As a result, instead of minimizing the cost function in (\ref{eq:joint-ML_total}) over all possible pairs $(s_1,s_2)$, we first obtain the estimate of $s_1$ using threshold detector, called $s_1^{ML}(s_2^{m})$, and then compute (\ref{eq:joint-ML_total}) for ($s_1^{ML}(s_2^{m}),s_2^{m}$), for $m=1,2, \cdots, M$. The optimal solution can be obtained as \vspace{-10pt}
\begin{equation}
\hat{s}_2=\argmin_{m} f\Big(s_1^{ML}(s_2^{m}),s_2^m\Big),
\label{eq:joint-ML-1_cond}
\end{equation}
where
\begin{align}
f\Big(s_1^{ML}(s_2^{m}),&s_2^m\Big) = \;|y_1 -   \frac{1}{\sqrt{\nu}} h_{1,1} \,g_{1,1} \,(s_1^{ML}(s_2^{m}) + \beta_1 s_2^m) 
\nonumber
\\
&-   \frac{1}{\sqrt{\nu}} h_{1,2} \,g_{1,2} \,(s_1^{ML}(s_2^{m}) + \beta_2 s_2^m)|^2\nonumber
\\&\,\,+|y_2-  \frac{1}{\sqrt{\nu}} h_{1,2} \,g_{1,2}(s_1^{ML}(s_2^{m}) + \beta_1 s_2^m) 
\nonumber
\\
&
-   \frac{1}{\sqrt{\nu}} h_{2,2} \,g_{2,2} \,(s_1^{ML}(s_2^{m}) + \beta_2 s_2^m)|^2.\label{eq:costfunc1}
\end{align}
Using the above described conditional ML decoding, we reduce the ML detection complexity of the proposed code  from ${\cal O}(M^2)$ to ${\cal O}(M)$ (see Algorithm \ref{Alg1}).
\begin{algorithm}
\caption{Conditional ML Decoding}
\label{Alg1}
{\bf Step 1:} Select $s_2^m$ from the signal constellation set. \newline 
{\bf Step 2:} Compute ${\tilde r} = r_1 + r_2$. \newline  
{\bf Step 3:} Supply ${\tilde r}$ into a phase threshold detector to get the estimate of $s_1$ conditional on $s_2^m$, called $s_1^{ML}(s_2^{m})$. \newline 
{\bf Step 4:} Compute the cost function in (\ref{eq:costfunc1}) for $s_1^{ML}(s_2^{m})$ and $s_2^{m}$.\newline
{\bf Step 5:} Repeat Step 1 to Step 4 for all the remaining constellation points.\newline
{\bf Step 6:} The $s_1^{ML}(s_2^{m})$ and $s_2^{m}$ corresponding to cost function with minimum value will be the estimate of $s_1$ and $s_2$.  \newline
\end{algorithm}
\subsection{Decoding Complexity Analysis}
In this section, we compare the computational complexity of the conditional ML decoding with that of the traditional ML decoding.
A simple measure to rate the complexity of any receiver is the number of complex Euclidean distances to compute. This is approximately proportional to the number of multiplications, which is generally more process intensive than additions. In Table~\ref{table:MLcomputationalcomplexity}, we summarize the number of arithmetic operations required by the traditional and conditional ML detectors for a $2 \times 2$ MIMO system with the signal constellation of size $M$.

\section{Simulation Results}
\label{sec:results}
In this section, we present the results of our numerical simulation to demonstrate the performance of the proposed coding scheme and compare it to the existing rate-two methods in the literature. In particular, we compare the BER performance of the proposed code with the \textit{VBLAST} \cite{foschini1996layered}, \textit{Matrix C} \cite{IEEEStd2006Air}, and \textit{MTD} \cite{rabiei2009new} schemes. Throughout the simulations, we assume a $2 \times 2$ MIMO structure and use 4-QAM constellation for symbol transmissions. We consider Rician fading channel model with the following form
\begin{align}
{\bf H}= \sqrt{\frac{K}{K+1}} {\bf H}_{L} + \sqrt{\frac{1}{K+1}} {\bf H}_w,
\label{eq:Rician_ch}
\end{align}
where $K$ is the Rician $K$-factor expressing the ratio of powers of the free-space signal and the scattered waves. Using this model, ${\bf H}$ is decomposed into the sum of a random component matrix, ${\bf H}_w$ and the deterministic component ${\bf H}_{L}$. The former accounts for the scattered signals with its entries being modeled as independent and identically distributed (i.i.d) complex Gaussian random variables with zero mean and unit variance. The latter, ${\bf H}_{L}$ models the LoS signals. In our simulations, the entries of matrix ${\bf H}_{L}$ are all set to one. This choice is motivated by the fact that optimal LoS MIMO channels are highly dependent on distance between the transmitter and receiver, and the antenna spacing~\cite{article_MIMO_LoS}. These conditions cannot be easily satisfied in mobile cellular networks. Hence, here, we have considered an ill-condition LoS channel. 

\begin{figure}
\centering
\includegraphics [scale=0.47]{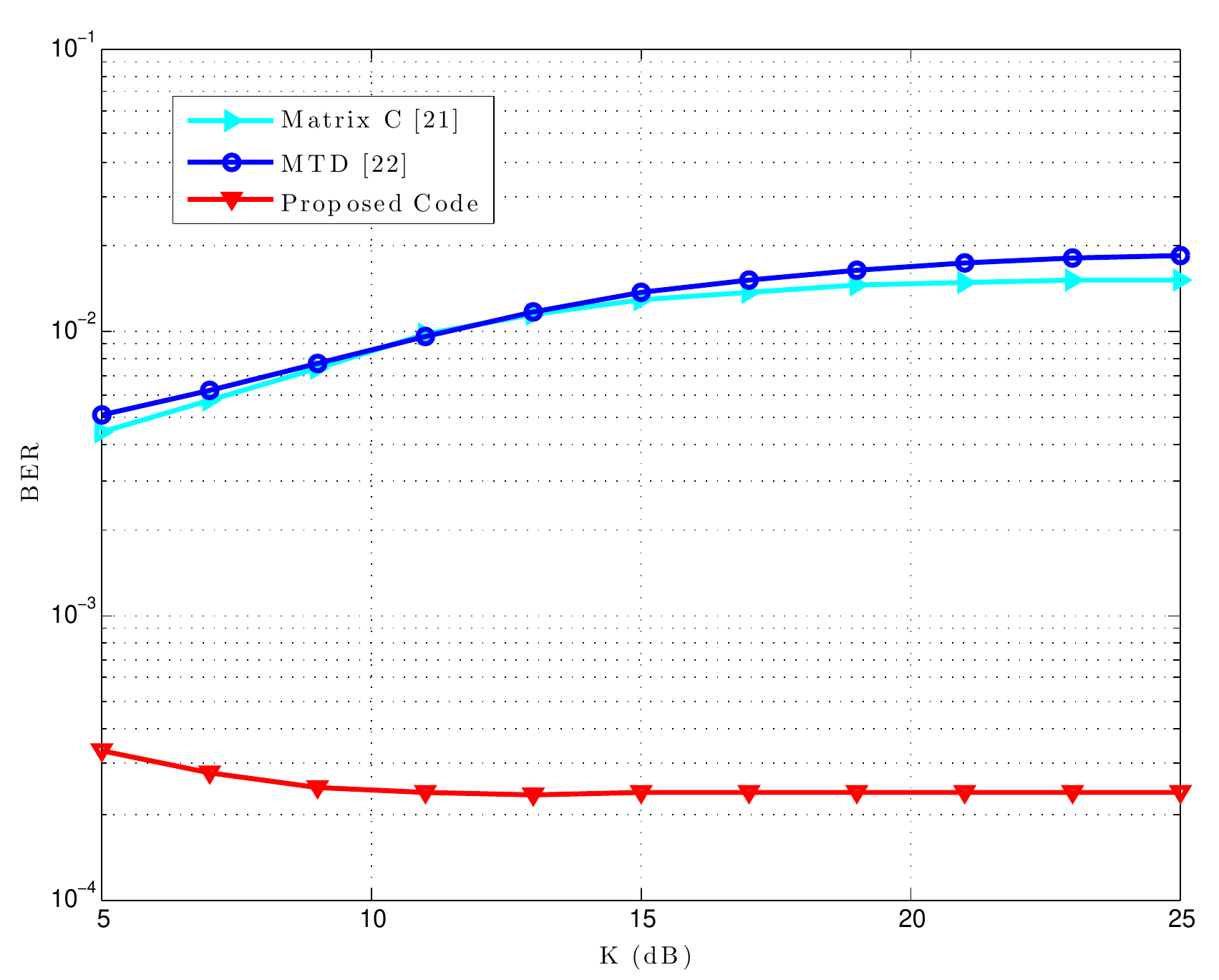}
\caption{\small BER performance of the proposed space code, the Matrix C, and MTD versus $K$, the Rician factor.}
\label{fig:OurCode_MTD_MatrixC_BERvsK}
\end{figure}

Fig.~\ref{fig:OurCode_MTD_MatrixC_BERvsK} shows the BER of the proposed space code, the Matrix C, and MTD versus $K$, the Rician factor. The BER of Matrix C and MTD degrades as $K$ increases, since as $K\rightarrow \infty$, the random component of the channel vanishes. Consequently, the channels reduces to ${\bf H}_L$. Under this condition, the channel becomes ill-condition as its covariance is low-rank. However, by reconfiguring the radiation pattern of each transmit-antenna pair, the proposed space codes can maintain a full-rank channel even when $K\rightarrow \infty$. Hence, as shown in Fig.~\ref{fig:OurCode_MTD_MatrixC_BERvsK} the BER performance of the proposed code remains invariant respect to changes in $K$, which is a key advantage of the proposed scheme for mm-wave applications. 

\begin{figure}
\centering
\includegraphics [scale=0.47]{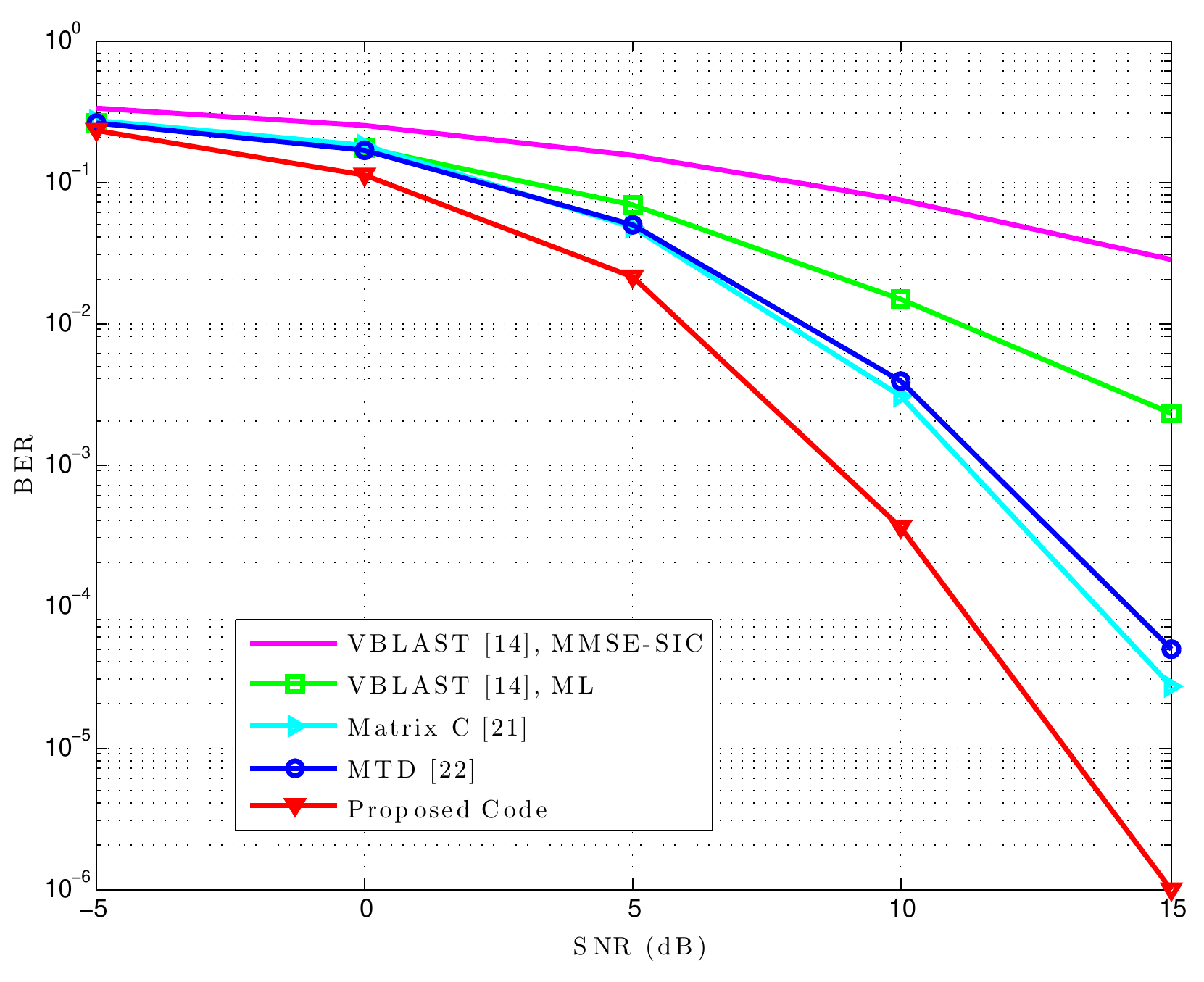}
\caption{\small BER performance of the proposed code with spectral efficiency of 4 bits per time slot in Rician fading channels.}
\label{fig:ber-performance-4bcu-Rician}
\vspace{-10pt}
\end{figure}

Fig. \ref{fig:ber-performance-4bcu-Rician} illustrates the BER performance of the proposed code in comparison with the performance of the VBLAST scheme and the aforementioned rate-two STBCs over a Rician fading channel with K-factor equal to $2$ dB. It can be observed from this figure that the proposed code outperforms all the considered codes. The second best performing code in this graph is Matrix C, which has been already included in the IEEE 802.16e-2005 specification \cite{sezginer2007full}. As this result indicates, at a BER of $10^{-4}$, the performance advantage of the proposed code compared with that of Matrix C is about $2.5$ dB. It also can be seen from this figure that at a BER of $10^{-3}$ the proposed code achieves more than 7 dB gain compared to the VBLAST scheme with ML decoding. In Table \ref{table:MLcomplexity}, we compare the ML decoding complexity of the proposed code with those of Matrix C, MTD, VBLAST for a $2 \times 2$ MIMO system. As shown in this table, the decoding complexity of the proposed code is ${\cal O}(M)$ which is substantially lower than the other codes.


\begin{table}
\caption{\small Computational complexity comparison.}
\begin{center}
     \begin{tabular}{|  p{3.5cm} |  p{3.5cm} |} \hline
Traditional ML Decoding & Conditional ML Decoding  \\ \hline
    \[ 
\left (
  \begin{tabular}{c}
  8 Multiplications  \\
  4 Subtractions  \\
  5 additions \\
  2 Squares 
  \end{tabular}
\right )\times M^2 
\]    $(M^2-1)$ Comparisons      &  \[ 
\left (
  \begin{tabular}{c}
  8 Multiplications  \\
  4 Subtractions  \\
  5 additions \\
  2 Squares 
  \end{tabular}
\right )\times M
\] $(M^2-1)$ Comparisons   \\ \hline
     \end{tabular}
\end{center}
\label{table:MLcomputationalcomplexity}
\end{table}
\begin{table}
\caption{\small Comparison of coding rate and ML decoding complexity.}
\begin{center}
	\begin{tabular}{| l | c | c |}
     \hline
     Coding Scheme & Symbol rate ($r_s$) & Complexity  \\ \hline
     Proposed code & 2 & ${\cal O}(M)$  \\ \hline
     Matrix C \cite{IEEEStd2006Air} & 2 & ${\cal O}(M^4)$  \\ \hline
     MTD  \cite{rabiei2009new} & 2 & ${\cal O}(M^2)$ \\ \hline
     VBLAST \cite{foschini1996layered} & 2 &${\cal O}(M^2)$  \\
     \hline
     \end{tabular}
\end{center}
\label{table:MLcomplexity}
\vspace{-10pt}
\end{table}

\section{Conclusions}
\label{sec:conc}
We proposed a rate-two space code for wireless systems employing reconfigurable antennas. It is indicated that such a setup can be advantageous for mmWave systems, since it can allow for the LoS-MIMO systems deployed in this band to achieve both spatial diversity and multiplexing. Moreover due to the structure of the proposed code and reconfigurable feature of the antenna elements, we reduce the ML detection complexity to be ${\cal O}(M)$, which has significant impact on the energy consumption of the receiver especially for higher order modulation schemes. We provided simulation results that demonstrate the performance of the proposed code and made comparisons with that of the previous coding schemes. As our results indicate the BER performance of the proposed code outperforms the rate-two STBCs and VBLAST scheme. However, it is important to consider that for future work, channel/directional of arrival estimation errors, phase noise, amplifier nonlinearity, and other issues pertaining to mmWave systems must also be considered to fully determine the potential of such $2\times 2$ MIMO systems in this band.


\bibliographystyle{IEEEtran}
\bibliography{IEEEabrv,Reference}

\end{document}